\documentclass[prd,preprint,eqsecnum,nofootinbib,amsmath,amssymb,preprintnumbers,tightenlines,longbibliography,hyperlink]{revtex4-1}

\usepackage{amsmath,graphicx}
\usepackage{color}
\usepackage[dvipsnames]{xcolor}
\usepackage[colorlinks,linkcolor=blue,citecolor=blue]{hyperref}

\definecolor{mygray}{gray}{0.6}

\def\x{{\boldsymbol x}}

\def\P{{\mathcal P}}

\def\L{{\mathcal L}}
\def\I{{\mathcal I}}

\def\R{{\mathcal R}}

\def\em{{\rm EM}}

\def\eq{{\rm eq}}
\def\p{{\boldsymbol p}}

\def\Eq#1{Eq.~(\ref{#1})}

\def\App#1{Appendix~\ref{#1}}
\def\Fig#1{Fig.~\ref{#1}}
\def\Sect#1{Section~\ref{#1}}
\def\Reff#1{Ref.~\cite{#1}}

\def\bra{\langle}
\def\ket{\rangle}
\def\bbra{\langle\!\langle}
\def\kket{\rangle\!\rangle}

\def\eg{{\it e.g.}}

\newcommand \beq{\begin{eqnarray}}
\newcommand \eeq{\end{eqnarray}}
\newcommand \be{\begin{equation}}
\newcommand \ee{\end{equation}}

\begin{document}

\title{Estimate magnetic field strength in heavy-ion collisions via the direct photon elliptic flow}

\author{Jing-An Sun}
\affiliation{Institute of Modern Physics, Fudan University, Handan Road 220, Yangpu District, Shanghai,
200433, China}

\author{Li Yan}
\affiliation{Institute of Modern Physics, Fudan University, Handan Road 220, Yangpu District, Shanghai,
200433, China}
\affiliation{Key Laboratory of Nuclear Physics and Ion-Beam Application (MOE), Fudan University, Shanghai 200433, China}

\date{\today}

\begin{abstract}

There must be electromagnetic fields created during high-energy heavy-ion collisions. As the quark-gluon plasma (QGP) starts to evolve hydrodynamically, although these fields may become weak comparing to the energy scales of the strong interaction, they are potentially important to some electromagnetic probes.  
In this work, we focus on the dissipative corrections in QGP due to the presence of a weak external magnetic field, and calculate accordingly the induced photon radiation in the framework of viscous hydrodynamics. By event-by-event hydrodynamical simulations, the experimentally measured direct photon elliptic flow can be well reproduced. Correspondingly, the direct photon elliptic flow implies a magnetic 
field strength around 0.1$m_\pi^2 \sim 10^{16}$ G. This is indeed a weak field in heavy-ion physics that is compatible to the theoretical predictions, however, 
it is still an ultra-strong magnetic field in nature.

\end{abstract}
\maketitle

\section{introduction}

One of the most significant achievements in high-energy heavy-ion experiments carried out at the Relativistic Heavy-Ion Collider at Brookhaven National Lab, and the Large Hadron Collider at CERN, is the discovery of a deconfined QCD matter -- Quark-Gluon Plasma (QGP)~\cite{RevModPhys.89.035001}. 

QGP has now been recognized as a perfect fluid, with small dissipations due to shear and bulk viscous effects. 
Both effects are related to the conservation of energy and momentum, which determines the three dimensional expansion of the QGP system. 
Shear viscous effect in QGP can be identified through the measurements of collective flow, observables that are used to quantify the momentum anisotropy of the emitted hadrons from high-energy nucleus-nucleus collisions~\cite{Heinz:2013th}. As the QGP  expands, its fluid nature would allow for a conversion between the initial spatial geometry and final state momentum spectrum~\cite{Ollitrault:1992bk}. Shear force, which presents 
as a fictional force
via the coupling between shear viscosity and gradients of hydrodynamic fields, e.g., $\eta\nabla^\mu u^\nu$, on the other hand, reduces the conversion ability, and hence suppresses collective flow~\cite{Teaney:2003kp,shear-Romatschke}.  
With respect to a large amount of experimental data on hadron collective flow, and comparisons to the event-by-event simulations based upon hydrodynamic modeling, the ratio of shear viscosity to entropy density, $\eta/s$, has been estimated
~\cite{Luzum:2008cw}.  Similarly, 
the bulk viscous corrections in QGP can be revealed from, for instance, the mean transverse momentum 
of the observed hadrons, which gets reduced with respect to a finite bulk pressure $\zeta \nabla^\mu u_\mu$~\cite{Ryu:2015vwa}.
With shear and bulk viscous corrections implemented, the hydrodynamic modeling of heavy-ion collisions has been successfully applied to the QGP evolution~\cite{Shen:2020mgh}, giving rise to theoretical characterizations of  
signatures related to the hadron spectrum 
in a huge parameter space including hadron species, momentum and harmonic orders~\cite{Bernhard:2016tnd,Bernhard:2019bmu}.

QGP is dissipative as well 
with respect to the transport phenomena associated with conserved currents.  As a medium system which is fundamentally dictated by strong (QCD) and electromagnetic (QED) interactions, in QGP one has the conservation of baryon charge and electric charge.\footnote{
We ignore the conservation of strangeness and charm in the current discussions.
} These conserved charges are expected substantial for low energies collisions. For instance, baryon charge conservation plays a key role in the search of QCD critical end point in the RHIC beam energy scan program (cf.~\Reff{Luo:2017faz} and references therein). However, for high energy nucleus-nucleus collisions, neither of them has been seriously taken into account. Neglecting baryon charge current appears reasonable at the top RHIC energy, considering the smallness of the ratio of net baryon charge over the charged particle multiplicity, $N_{\rm baryon}/N_{\rm ch}\sim 10^{-4}$~\cite{STAR:2014egu}. The electric current, on the other hand, can result in potentially sizable effect as it is related to the electromagnetic forces exert on the QGP medium, $\sigma_{\rm el} \vec E$. 
Here, $\sigma_{\rm el}$ is the electrical conductivity.

For non-central collisions, with respect to the space-time configuration of the colliding nuclei, there {\it must} be finite electromagnetic fields created in the collision region. At the instant when the two nuclei collide, one expects a net magnetic field pointing out of the reaction plane, with its magnitude $|eB|/m_\pi^2\sim 10$ at the top RHIC energy and $|eB|/m_\pi^2\sim 10^2$ at the LHC, with $m_\pi$ the pion mass~\cite{McLerran:2013hla,Skokov:2009qp,Bzdak:2011yy,Deng:2012pc}. Albeit strong initially, as the system evolves, the magnetic field decays drastically, which scales as $1/ t^3$.  Unless the QGP is sufficiently conductive in its pre-equilbrium stage, the external magnetic field should become weak
when the medium starts to evolve hydrodynamically~\cite{Yan:2021zjc,Stewart:2021mjz,Huang:2022qdn}. In fact, theoretical analysis found that the field can drop to $10^{-2} m_\pi^2\sim 10^{15}$ G in the QGP system in less than 1 fm/c, which is three orders of magnitudes smaller comparing to its initial value. However, let us emphasize that although it is a weak field relative to the energy scale in the QGP medium, \eg, 
temperature sqare in QGP in heavy-ion experiments can range from $\sim 1.2m_\pi^2$ to $\sim 10 m_\pi^2$,
it is still an ultra-strong magnetic field in nature. For instance, the strongest magnetic field observed so far in neutron star reaches $1.6\times 10^{13}$ G~\cite{Kong:2022cbk}.

In this article, we would like to 
show that, the weak magnetic field during the QGP evolution, and accordingly the associated dissipative corrections, can be visible through the observed electromagnetic probes. In particular, we focus on the photon radiation from the QGP in the presence of a weak external magnetic field. Following the formulation developed in \Reff{Sun:2023pil}, with even more realistic event-by-event simulations, we find that the weak magnetic field 
 is not only responsible to the disagreement between theory and experiments on the direct photon elliptic flow~\cite{PHENIX:2015igl,ALICE:2018dti}, but also provides an extra source to the direct photon triangular flow.
 As a consequence of its sensitivity, we propose to take the experimentally measured elliptic flow of the direct photons as an ideal probe to the magnetic field in QGP.
 
 The paper is organized as follows. In \Sect{sec:wmpe}, we first give a brief review on the derivation of dissipative corrections to photon production in QGP, with respect to 
 shear and bulk viscous effects, using the Chapman-Enskog method. As a generalization of \Reff{Sun:2023pil}, we then present the detailed formulation for photon production in QGP in the presence of a weak external magnetic field. In \Sect{sec:num}, with the novel formulation implemented, we calculate the weak magnetic field induced photon production in QGP via the event-by-event hydrodynamic simulations. With respect to the experimental data on direct photon elliptic flow, we give an estimate on the magnetic field at the top RHIC energy, in \Sect{sec:bfield}. Conclusion and discussions will be given in \Sect{sec:dis}.

%


\section{Off-equilibrium photon emission from QGP}
\label{sec:wmpe}

Thermal radiations of photons from QGP  out of local equilibrium can be calculated in a kinetic theory approach. Given the photon phase space distribution function, $f_\gamma(X, \p)$,
the evolution of photon density in phase space satisfies~\cite{Arnold:2001ms},\footnote{
Throughout the paper, we use capital letters for four-vectors, while bold lower case letters are used as normal three vectors, e.g., $P^\mu=(E_p, {\bf p})$. We take the most negative matrix convention, $g_{\mu\nu} = {\rm diag}(+,-,-,-).$
}
\be
\frac{1}{(2\pi)^3}
P^\mu \partial_\mu f_\gamma(X, \p) = 
E_p\frac{d^3\R}{d^3 \p} [1 + f_\gamma(X,\p)] 
- \mbox{absorption}\,.
\ee
The first term on the right-hand side of the equation originates from photon emission, with the differential rate $E_p d^3 \R/d^3\p$ characterizing photon productions per unit space-time volume, while the second term describes the absorption of photons. Note that absorption is related to emission via the detailed balance condition. For a QGP medium in which the mean free path of photons is much longer than the system size, photon absorption can be neglected. Therefore, for the 2-to-2 scattering processes in QGP, for instance, the photon emission rate per unit volume can be evaluated via the collision integral. In terms of the phase space distribution functions of quarks, antiquarks and gluons in the medium, it is,
\begin{align}
\label{eq:rate}
E_p \frac{d^3 \R}{d^3 \p} =
\frac{g}{2(2\pi)^3}\sum_{i}& \int\frac{d^3 \p_1}{(2\pi)^3 2 E_{1}}\frac{d^3 \p_2}{(2\pi)^3 2 E_{2}}\frac{d^3 \p_3}{(2\pi)^3 2 E_{3}} |{\cal{M}}_i|^2 f_1(P_1) f_2(P_2)(1\pm f_3(P_3))  \cr
&\times   (2\pi)^4 \delta^{(4)} (P_1+P_2-P_3-P)\,,
\end{align}
where $g$ takes into account spin and color degrees of freedom, %
$f_1$, $f_2$ and $f_3$ are phase space distribution functions of gluons and quarks (antiquarks), respectively. The
summation with respect to the scattering amplitude $|{\cal{M}}_i|$ is taken over quark-antiquark annhiliation and Compton scattering processes~\cite{Kapusta:2006pm,Arnold:2001ms}. 
       

For photons of energy scale much greater than the medium temperature, $E_p\gg T$, 2-to-2 elastic processes are expected dominated by small angle scatterings. Accordingly, the differential rate in \Eq{eq:rate} can be simplified~\cite{Churchill:2020uvk,Berges:2017eom},
\be
\label{eq:rate_sg}
E_p \frac{d^3 \R}{d^3 \p} \approx
C \alpha_{\rm EM} \alpha_s\I \L_c  (f_q + f_{\bar q})\,,
\ee
with $C$ a constant resulted from the summation over quark spin, color and flavor. 
Here, $\L_c$ is a Coulomb logarithm that relies on the separation of hard and soft energy scales in the medium, for instance, $\L_c \sim \log \alpha_s^{-1}$. In practice, its explicit value can be adjusted to match the analytical results of the differential rate as a function of temperature. In the small angle approximation, effect of the QGP medium has been absorbed into the constant $\I$~\cite{Blaizot:2014jna},
\be
\I = \int \frac{d^3 \p}{(2\pi)^3 E_p}(f_g + f_q + f_{\bar q})\,,
\ee  
which, together with the product of the electromagnetic and strong coupling constants $\alpha_{\rm EM}\alpha_s$, effectively captures the ability of conversion between a quark (antiquark) and a photon inside  the QGP medium. 
\Eq{eq:rate_sg} simply relates the photon spectrum to the phase space distribution of quarks and anti-quarks, from which, one expects anisotropic photon emission from QGP as long as quarks or anti-quarks in the medium exhibit elliptic and triangular flow signatures. This explains the origin of direct photon elliptic flow observed in viscous hydrodynamic simulations~\cite{Shen:2013cca,Gale:2014dfa,Gale:2021emg}.

The resulted differential rate \Eq{eq:rate} applies to medium in or out of equilibrium, provided that out-of-equilibrium effect does not affect the scattering amplitude substantially~\cite{Huang:2022fgq}.
For a QGP in local equilibrium, whose evolution is describable by ideal hydrodynamics, the equilibrium distributions are known as the Bose-Einstein distribution for gluons, 
\be
\label{eq:ng}
n_g(P) = \frac{1}{\exp{(P\cdot U/T)}-1}
\ee
and the Fermi-Dirac distribution for quarks (and antiquarks),
\be
\label{eq:nq}
n_q(P) = n_{\bar q}(P) =  \frac{1}{\exp(P\cdot U/T) + 1}\,.
\ee
The flow four velocity $U^\mu$ and temperature $T$ of the medium should be
 determined by a hydrodynamical modeling of the system evolution, 
accordingly thermal radiations from an equlibriated QGP are well determined. 
 In both \Eq{eq:ng} and \Eq{eq:nq}, we have neglected the effect of charge conservation, so that chemical potentials associated with the conserved charges vanish.  This is a good approximation for high energy heavy-ion collisions, where the net conserved charges (e.g., baryon, electric or strangeness, etc) are negligible. However, as the collision energy decreases as in the RHIC beam energy scan progrem, net charge density in the system increases~\cite{STAR:2014egu}, a finite chemical potential should be taken into account. Without the effect of chemical potential, \Eq{eq:nq} implicitly assumes the identification of distribution functions between quarks and antiquarks. However, deviations could arise, for instance, from the difference in flow velocities when external electromagnetic forces apply to the system evolution. 

{
Inelastic process of the photon emission from QGP, such as in-medium bremsstrahlung and inelastic pair annihilation, could be as important as the 2-to-2 elastic scatterings, due to the near-colinear singularities~\cite{Arnold:2001ms}. In the presence of an external magnetic field, there also exists 1-to-2 contribution, e.g., $q\to q\gamma$~\cite{Tuchin2013,Tuchin2015,wang2020,wang2021,zakharov2016effect,zakharov2016synchrotron}, as quarks receive extra momentum transfer from the magnetic field and become effectively off shell. 
In both cases, however, the produced photons dominate over quite different kinematic regions, comparing to the elastic scattering process considered in this study. The sum of the in-medium bremsstrahlung and inelastic pair annihilation give rise to the major source of photon production of large momentum~\cite{Arnold:2001ms}, while the magnetic field induced inelastic radiation favors photon radiated with small transverse momentum~\cite{wang2021}. Therefore, as most of the photons from experiments are of intermediate transverse momentum, where the elastic scatterings generate the largest contribution, for the moment we neglect these inelastic contributions.}

\subsection{Dissipations in QGP driven by spatial gradients}

When a QGP system is driven slightly out of local equilibrium, its evolution can be described by dissipative fluid dynamics. Correspondingly, the phase space distributions of constituents {\it must} contain small dissipative corrections, which can be solved in a perturbative manner by means of the Chapman-Enskog approximation~\cite{DeGroot:1980dk}. 

As an illustration, let us first consider
the dissipative effects in QGP driven by spatial gradients $\nabla$, which in hydrodynamics leads to a systematic expansion of the stress tensor,
\be
\Pi^{\mu\nu} \equiv \pi^{\mu\nu} - \Pi \Delta^{\mu\nu} = 2 \eta \bra \nabla^\mu U^\nu \ket - \zeta \Delta^{\mu\nu} \nabla\cdot U + O(\nabla^2)\,,
\ee
where the projection operator is $\Delta^{\mu\nu}=U^\mu U^\nu - g^{\mu\nu}$ and $\nabla^\mu = \Delta^{\mu\nu}\partial_\nu$. Here and in what follows, the angular bracket around a tensor is used to indicate it being symmetric, traceless and transverse to the flow four velocity. 
At the leading order of the expansion, shear and bulk viscous corrections emerge and one has the Navier-Stokes hydrodynamics. In phase space distributions, the gradient expansion leads to corrections as well.
With respect to the Boltzmann equation in a relaxation time approximation, the distribution function $f_a$ (with $a=$ parton species) satisfies
\be
\label{eq:rta}
P^\mu \partial_\mu f_a = - \frac{P\cdot U}{\tau_R} \delta f_a\,,
\ee
where $\delta f_a \equiv f_a - n_{a,{\rm eq}}$ characterizes the deviation of the phase space distribution from local equilibrium. In the relaxation time approximation, scatterings among quarks and gluons are captured by the scalar parameter $\tau_R$. Effectively, $\tau_R$ can be evaluated at the linearised order of the collision integral. In general, $\tau_R$ depends on the medium properties such as temperature, $T$, as well as the dynamical properties of the particle such as momentum, $P\cdot U$~\cite{Dusling:2009df}.  For a multi-component system, $\tau_R$ should also rely on constituent species. Throughout the current study, we shall ignore the momentum dependence in $\tau_R$, and the dependence on particle species, unless when it is necessary for discussions. In practice, as in hydrodynamics where the underlying dynamical properties are constrained via transport coefficients, $\tau_R$ can be also identified in terms of the transport coefficients, such as the shear viscosity $\eta$ and bulk viscosity $\zeta$. 

When $f_a$ is expanded in gradients,
\be
f_a= n_{a,{\rm eq}} + \delta f_a^{(1)} + \delta f_a^{(2)} + \ldots,
\ee
with the superscript introduced to label the expansion order with respect to spatial gradient, \Eq{eq:rta} can be solved order by order. 
The first order solution is
\begin{align}
\label{eq:df1}
\delta f_a^{(1)}(X, P) &= - \frac{\tau_R}{P \cdot U} P^\mu \partial_\mu n_{a,{\rm eq}}
+\delta n_{a,{\rm eq}}^*\cr
&= - \frac{\tau_R}{P \cdot U}\frac{n_{a,{\rm eq}}'}{T}\left\{
 P^\mu P^\nu \bra \nabla_\mu U_\nu\ket +  \left[(P\cdot U)^2\left(c_s^2- \frac{1}{3}\right)+ \frac{m_a^2}{3}\right]\nabla\cdot U
\right\}\cr
& \quad + n_{a,{\rm eq}}'\left(\frac{P^\mu \delta U_\mu^*}{T} - \frac{P\cdot U}{T^2} \delta T^*\right)
\end{align}
where $c_s = \sqrt{\partial P/\partial e}$ is the sound velocity and $m_a$ the mass of parton species $a$.  In \Eq{eq:df1} and in what follows, the prime indicates derivative of the equilibrium distribution with respect to $P\cdot U/T= E_p/T$, therefore, $n'_{a,{\rm eq}}=-n_{a,{\rm eq}}(1- s n_{a,{\rm eq}})$, with $s=1$ or $s=-1$ corresponding to quarks or gluons, respectively.
In principle, the gradient corrections affect the Landau's matching conditions order by order, which in turn gives rise to gradient corrections in flow four-velocity $\delta U^*_\mu$ and temperature $\delta T^*$. These are included in $\delta n_{a,\eq}^*$.
In obtaining \Eq{eq:df1}, 
we have used the thermodynamic relations, e.g., $Ts=e+P$, and
the equations of motion of hydrodynamics have been considered up to the order of $O(\nabla)$,
\be
(e+\P)D  U^\alpha = \nabla^\alpha \P + O(\nabla^2)\,,\qquad
D e = - (e + \P) \nabla\cdot U + O(\nabla^2)\,,
\ee
where $D = u^\mu \partial_\mu$.
Note that these equations can be interpreted as the acceleration of a fluid cell due to an external force driven by gradient, and the conservation of energy in the presence of gradients, respectively.

Distribution functions from kinetic theory and hydrodynamics are related with each other with respect to the Landau's matching conditions. For the stress tensor, one has
\begin{align}
\label{eq:matching}
\Pi^{\mu\nu} &= \sum_a \int \frac{d^3 \p}{(2\pi)^3 E_p}  P^\mu P^\nu  \delta f_a\,.
\end{align}
Substitute \Eq{eq:df1} into the matching condition, one finds 
\begin{align}
\label{eq:tauR_1st}
\eta 
 = \tau_R \chi_{\rm shear}\,,\qquad
\zeta 
= \tau_R \chi_{\rm bulk}\,,
\end{align}
where 
$\chi_{\rm shear}$ and $\chi_{\rm bulk}$ are effective susceptibilities defined through the correlations among components of the energy-moment tensor, with respect to the shear and bulk viscous corrections. For instance, $\chi_{\rm shear}\propto e+\P$ in the massless limit. More details on these quantities are given in \App{app1}. Given the relations in \Eq{eq:tauR_1st}, we are allowed rewrite the viscous corrections to the distribution function in terms of the shear and the bulk viscosities. For the shear viscous correction, one finds
\be
\label{eq:dfp}
\delta f_{a,\pi}^{(1)} 
= - \frac{\eta n_{a,{\rm eq}}'}{T P\cdot U \chi_{\rm shear}}P^\mu P^\nu \bra \nabla_\mu U_\nu\ket 
= -\frac{ n_{a,{\rm eq}}'}{2 T P\cdot U \chi_{\rm shear}}P^\mu P^\nu \pi_{\mu\nu} + O(\nabla^2) \,,
\ee 
while for the bulk viscous correction, one finds,
\begin{align}
\label{eq:dfpp}
\delta f_{a,{\rm \Pi}}^{(1)} & 
= -\frac{\zeta n_{a,{\rm eq}}'}{T \delta P\cdot U \chi_{\rm bulk}} \nabla\cdot U\left[(P\cdot U)^2\left(c_s^2- \frac{1}{3} - \frac{\xi_2}{\xi_4}\right)+ \frac{m_a^2}{3}\right] \cr
&=\frac{\Pi n_{a,{\rm eq}}'}{T P\cdot U \chi_{\rm bulk}} \left[(P\cdot U)^2\left(c_s^2- \frac{1}{3} - \frac{\xi_2}{\xi_4} \right)+ \frac{m_a^2}{3}\right]
+ O(\nabla^2)\,.
\end{align}
Note that the first order bulk viscous correction modifies the matching condition regarding temperature, with $\delta T^*\sim \nabla\cdot U$, resulting additionally a ratio between two integrals, $\xi_2/\xi_4$, as shown in \App{app1}. It can be seen that as the conformal limit is approached, either by $m_a\to 0$ or $c_s^2\to 1/3$, \Eq{eq:dfpp}, as well as $\chi_{\rm bulk}$, vanish. 

Since we are considering a constant relaxation time, in \Eq{eq:dfp} and \Eq{eq:dfpp}, an extra $P\cdot U$ factor appears in the denominator, which makes the expressions slightly different from those used in \Reff{Putschke:2019yrg}. This extra factor would be absorbed if the relxation time has a linear momentum dependence as $\tau_R \propto (P\cdot U)$, corresponding to the quadratic ansatz considered in \Reff{Dusling:2009df}. 
In both cases of the shear and the bulk viscous corrections, 
the substitution of $2\eta\bra \nabla_\mu U_\nu\ket$ and $\zeta\nabla\cdot U$ has been made in the expressions by the shear stress tensor $\pi^{\mu\nu}$ and bulk pressure $\Pi$, respectively, which are valid up to corrections of order of $\nabla^2$. The substitution is practically convenient since $\pi^{\mu\nu}$ and $\Pi$ are dynamical variables in the hydrodynamical modeling of QGP evolution~\cite{Heinz:2013th}, which are achievable from numerical simulations. \Eq{eq:dfp} and \Eq{eq:dfpp} have been widely applied, especially for the cases of thermal photon radiations in QGP considering viscous corrections~\cite{Gale:2021emg}.

\subsection{Dissipations in QGP due to the weak electromagnetic fields} 

In the presence of external electromagnetic fields, a medium with charge carriers can be driven out of local equilibrium owing to the electromagnetic forces. Accordingly, the conserved current receives dissipative corrections,
\be
j^\mu \equiv n_c U^\mu + N^\mu = n_c U^\mu + \sigma_{\rm el} E^\mu  + O(\nabla n_c) 
+ O(|eF^{\mu\nu}|^2)\,,
\ee
where $n_c=e \sum_a Q_a n_a$ is the net charge density and $Q_a$ is the electric charge number of constituent species $a$. The first order dissipation due to electromagnetic fields follows the standard Ohm's law, with electrical conductivity $\sigma_{\rm el}$ the corresponding transport coefficient and $E^\mu= F^{\mu\nu} U_\nu$ the electric field in the local rest frame of the fluid. In principle, there should be diffusion contributions to the current via the gradients of the net charge density, e.g., $\nabla^\mu n_c$, which we shall nevertheless neglect assuming that QGP created in high-energy heavy-ion collisions satisfies the condition of local charge neutrality, 
namely, $n_c \approx 0$.\footnote{
There can be contributions from thermal fluctuations to the net charge density, namely, non-vanishing contributions from the thermal ensemble average of multi-point corrrelators to local net charge density, such as $ \bbra n_c\kket\propto \bbra (\delta n_c )^2 \kket^{1/2} \propto \sqrt{T\sigma_{\rm el}}$. On an event-by-event basis, net charge density receives also contributions from initial state fluctuations, which approximately depends inversely on the square root of charged multiplicity, $n_c\sim 1/\sqrt{N_c}$.
}

Following the strategy for the shear and bulk viscous corrections, we now derive the dissipative correction to the distribution function of QGP in the presence of a weak external electromagnetic field. With respect to the weak field condition, $|eF^{\mu\nu}|\ll T^2$, we first notice that the distribution function now admits an expansion in terms of the field strength,
\be
\label{eq:faEM}
f_a = n_{a,{\rm eq}} + \delta f_{a,{\rm EM}}^{(1)} + \delta f_{a,{\rm EM}}^{(2)} + \ldots
\ee
where again the superscript labels expansion order. One expects that \Eq{eq:faEM} solves the Boltzmann equation, but now with a Vlasov term which takes into account the effect of external electromagnetic fields~\cite{DeGroot:1980dk},
\be
P^\mu \partial_\mu f_a + e Q_a  F^{\mu\nu}P_\mu\frac{\partial f_a}{\partial P^\nu} = - \frac{P\cdot U}{\tau_R}\delta f_a\,.
\ee
At the linearised order of the field strength, the solution can be found as, 
\be
\label{eq:fem1}
\delta f_{a,{\rm EM}}^{(1)}(X,\p) =-\frac{\tau_R}{P\cdot U} eQ_a F^{\mu\nu}P_\mu\frac{\partial n_{a,{\rm eq}}}{\partial P^\nu}= -eQ_aF^{\mu\nu}P_\mu U_\nu \frac{\tau_R}{P\cdot U} \frac{n_{a,{\rm eq}}'}{T} \,.
\ee
Apparently, the solution applies to quarks but not to gluons, since gluons are electrically charge neutral.



For a system with multi-component contributions to the charge carriers such as QGP, and ignore local net charge density, the charge current is related to the distribution function of charge carriers through the Landau matching condition, $
N^\mu = \sigma_{\rm el} E^\mu = e\sum_a Q_a  N^\mu_a 
$, where the dissipative current associated with species $a$ is,
\be
N_a^\mu = g_a \int \frac{d^3 \p}{(2\pi)^3 E_p} P^\mu \delta f_{a,{\rm EM}}^{(1)}
=e g_a Q_a \tau_R E^\mu \chi_a\,.
\ee
Here, the scalar function determined via the integral,
\be
\chi_a = -\frac{1}{3}\int \frac{d^3 \p}{(2\pi)^3 E_p} (P^\alpha P^\beta \Delta_{\alpha \beta} )\frac{n_{a,{\rm eq}}'}{P\cdot U}\,,
\ee
quantifies nothing but the electric charge susceptibility of parton species $a$. 
As a consequence, one finds the relation between the electric charge conductivity, relaxation time $\tau_R$, and the  electric charge susceptibility of QGP
$\chi_{\rm el}$,
\be
\sigma_{\rm el} = \tau_R \chi_{\rm el}\,,
\ee
where
\be
\chi_{\rm el}
=4\pi\alpha_{\rm EM}\sum_a g_a Q_a^2 \chi_{a,{\rm el}}\,.
\ee
Accordingly, given the relation one may replace $\tau_R$ by the electric conductivity, $\tau_R=\sigma_{\rm el}/\chi_{\rm el}$, so that the dissipative correction to the distribution function due to weak external electromagnetic fields becomes,
\be
\label{eq:fem1b}
\delta f_{a,{\rm EM}}^{(1)}(X,\p)= -\frac{n_{a,{\rm eq}}'}{P\cdot U} \frac{\sigma_{\rm el}}{ T \chi_{\rm el}} e Q_a F^{\mu\nu} P_\mu U_\nu  = -\frac{n_{a,{\rm eq}}'}{T \chi_{\rm el} P\cdot U} e Q_a N^\mu P_\mu\,.
\ee
The first expression has been taken into account in \Reff{Sun:2023pil}, while the second equation facilitates numerical applications when the charge current $N^\mu$ becomes a dynamical variable in simulations. Especially, with a proper formulation of the conserved current, the second expression applies also to cases with a finite local charge density of electric charge, baryon charge, etc. 
Generalization to more complicated situations in which relaxation time differs among particle species, such as QGP with thermalized heavy quark components, or hadron gas in which hadron masses can be order of magnitudes different, is straightforward.


There are a couple of comments in order. First, in deriving \Eq{eq:fem1}, as expected in the Chapman-Enskog procedure, there exist extra contributions from the equations of motion of hydrodynamics up to order $O(|eF^{\mu\nu}|)$. Indeed, one has
\be
(e+\P) D U^\mu =  E^\mu n_c  +O(|eF^{\mu\nu}|^2)\,,
\ee 
which characterizes the acceleration of a charged fluid cell due to the external electromagnetic forces. Accordingly, 
from the Boltzmann-Vlasov equation one finds, in addition to \Eq{eq:fem1}
\be
\label{eq:fem1a}
\delta f_{a,{\rm EM}}^{(1)}(X,\p) \supset
-\frac{\tau_R}{P\cdot U}P^\mu \partial_\mu n_{a,{\rm eq}} = n_c \frac{n_{a,{\rm eq}}'}{T}\frac{\tau_R}{e+\P}  F^{\mu\nu}P_\mu U_\nu
\ee
However, again, we neglect this contribution as in QGP created in high energy collisions, local net charge density is negligible, $n_c \approx 0$. Of course, as the collision energy decreases, acceleration due to the external electromagnetic field can be substantial, not only because of a finite local net charge density in the medium, but also due to the fact that the external electromagnetic fields become stronger. Secondly, as a perturbation around local equilibrium, \Eq{eq:fem1} implies that 
\be
\label{eq:smallfem}
\frac{\delta f_{a,{\rm EM}}^{(1)} }{n_{a,{\rm eq}}}\sim \frac{e F^{\mu\nu}}{T^2}\times( \tau_R T) 
\sim \frac{|e F^{\mu\nu}|}{T^2} \frac{\sigma_{\rm el}}{T\gamma}\ll 1\,.
\ee
The factor linear in the field strength can be recognized small by the condition of weak external field. The relation is further constrained by the appearance of the relaxation time $\tau_R$. Since $\tau_R^{-1}$ captures the collisions, \Eq{eq:smallfem} re-interprets the fact that the forces induced by external electromagnetic fields are sub-leading in comparison to collisions among quarks and gluons. Therefore, 
the system should always stay close to local equilibrium irrespective of the presence of external fields. The condition \Eq{eq:smallfem} validates the Chapman-Enskog method for solving the Boltzmann-Vlasov equation.

\subsection{Weak magnetic photon emission}

\begin{figure}
\begin{center}
\includegraphics[width=0.3\textwidth] {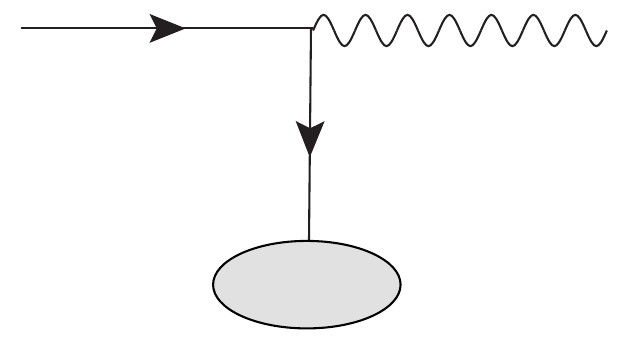}
\hspace{20mm}
\includegraphics[width=0.3\textwidth] {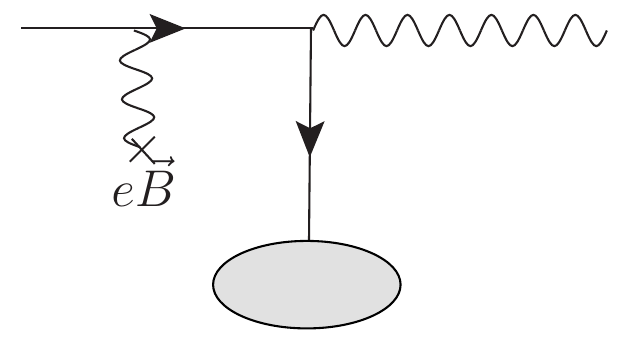}
\caption{
\label{fig:gammaEB}
Diagrammatic illustration of the photon emission without and with weak external magnetic fields from QGP in the small angle approximation. The blobs represent the effect of QGP medium that has been integrated and absorbed into the function $\mathcal{I}$.
}
\end{center}
\end{figure}

With respect to the dissipative corrections to the quark distributions, Eqs~(\ref{eq:dfp}), (\ref{eq:dfpp}) and (\ref{eq:fem1b}),  photon thermal radiations from a viscous QGP in the presence of a weak external electromagnetic field can be calculated. Up to the linearized order of spatial gradient and external field strength, the emission rate now contains dissipative contributions,
\be
E_p \frac{d \R}{d^3 \p}\equiv E_p \frac{d \overline{\R}}{d^3 \p}+E_p \frac{d \R^{\rm EM}}{d^3 \p} \approx C \alpha_s \alpha_{\rm EM}\I \L_c\sum_a (n_{a,{\rm eq}} + \delta f_{a,\pi}^{(1)} + \delta f_{a,\Pi}^{(1)} + \delta f_{a,{\rm EM}}^{(1)})\,,
\ee
where for later convenience, we have identified separately, $E_p d \overline{\R}/d^3 \p $
as the thermal radiation from a background system without the influence of external electromagnetic fields, and
\begin{align}
\label{eq:rateem}
E_p \frac{d \R^{\rm EM}}{d^3 \p} \approx C \alpha_s \alpha_{\rm EM}\I \L_c\sum_a \delta f_{a,{\rm EM}}^{(1)}
= -C \alpha_s\alpha_{\rm EM} \I \L_c \sum_a \frac{n_{a,{\rm eq}}'}{P\cdot U} \frac{\sigma_{\rm el}}{T \chi_{\rm el}} e Q_a F^{\mu\nu} P_\mu U_\nu\,,
\end{align}
as the thermal photon emission from QGP induced entirely by the weak external fields. These separated sources of photon emissions are illustrated in \Fig{fig:gammaEB}, which depicts respectively the conversion of a quark and a quark inside external magnetic field to a photon in the small angle approximation.  Under the weak field condition $|eF^{\mu\nu}|\ll T^2$, the influence from the weak magnetic field on the quark propagator and vertex can be negligible~\cite{Huang:2022fgq}, while the conversion ability, which is characterized by the constant $\I$ (blobs in \Fig{fig:gammaEB}), is not affected by the external magnetic field, since,
$
\delta \I = \int \frac{d^3 \p}{(2\pi)^3 E_p} \delta f_{a,{\rm EM}}^{(1)}  = 0 \,.
$

After a space-time integral with respect to the QGP expansion, the emission rate leads to the radiated thermal photon spectrum, for instance, the background contribution leads to photon spectrum
\be
\label{eq:specbkg}
E_p \frac{d^3 \overline{ N}}{d^3 \p} =\int d^4 X \left(E_p \frac{d \overline{\R}}{d^3 \p} \right) = 
\bar v_0\left[1 + \sum_{n=1} 2 \bar v_n \cos n(\phi_p - \Psi_n)\right]\,,
\ee
and similarly, the photon emission by the weak electromagnetic fields has the spectrum
\be
\label{eq:specem}
E_p \frac{d^3  N_{\rm EM}}{d^3 \p} = \int d^4 X \left(E_p \frac{d \R^{\rm EM}}{d^3 \p} \right) = 
v_0^\em \left[1 + \sum_{n=1} 2 v_n^\em \cos n (\phi_p - \Psi_n)\right] \,,
\ee
where $\phi_p$ is the azimuthal angle of the photon, $\Psi_n$ are the reference planes of the n-th order flow harmonics determined by charged particles.
In practical simulations as in the current work, the photon emissions from the background system can be generalized to include all possible sources. 
For instance, for the calculation of direct photons in $E_pd^3\overline{N}/d^3 \p$ there should be photons from prompt hard scatterings, QGP thermal radiations from elastic and inelastic scatterings, as well as thermal radiations from a hadron gas. 
In both \Eq{eq:specbkg} and \Eq{eq:specem}, coefficients of the expansion into harmonics refer to respectively as photon yields with $n=0$, elliptic flow with $n=2$ and triangular flow with $n=3$, etc.. These are experimental measurables that characterize the emission anisotropy of photons. Taken both sources into account, one expects the observed direct photon spectrum 
\be
E_p \frac{d^3 N^\gamma}{d^3 \p} = E_p \frac{d^3 \overline{ N}}{d^3 \p}+E_p \frac{d^3  N_{\rm EM}}{d^3 \p} 
= v_0^\gamma(1+ 2 v_2^\gamma \cos 2(\phi_p - \Psi_2) + 2 v_3^\gamma \cos 3(\phi_p - \Psi_3) + \ldots)\,,
\ee
with in particular, 
\be
v_0^\gamma = \bar v_0 + v_0^\em\,,\qquad
v_2^\gamma = \frac{\bar v_0 \bar v_2 + v_0^\em v_2^\em}{\bar v_0 + v_0^\em}\,,\qquad
v_3^\gamma = \frac{\bar v_0 \bar v_3 + v_0^\em v_3^\em}{\bar v_0 + v_0^\em}\,,
\ee
the yields, elliptic flow and triangular flow of the direct photons, respectively.

The radiated photon spectrum from the background medium has been analyzed extensively, especially in the framework of viscous hydrodynamics~\cite{Paquet:2015lta}. The resulted emission anisotropy, e.g., $\bar v_2$ and $\bar v_3$, are mostly associated with the medium expansion with respect to initial state geometries. For the radiated photons induced by external electromagnetic fields, on the other hand, the spectrum is highly anisotropic as a consequence of the interplay between the external electromagnetic fields and the dynamics of the background medium expansion~\cite{Sun:2023pil}. In realistic heavy-ion collisions, space-time configuration of the colliding nuclei demands that the external fields are well oriented,
$\vec B = B_y \hat y$ and $\vec E = E_x \hat x$, both of which contribute one dipole mode, $\cos\phi_p$, to the rate in \Eq{eq:rateem}, as well as to the weak field emitted photon spectrum, in \Eq{eq:specem}. Provided that the background QGP exhibits azimuthal anistropies, with non-vanishing $\cos n\phi_p$ terms in the background quarks ditributions, and in particular with a dipole moment which is related to charged hadron $v_1$, one finds a $\cos 2\phi_p$ mode which gives rise to photon elliptic flow, $v_2^\em$. In principle, $v_2^\em$ alone can be remarkably significant, which leads to a finite increase in the observed direct photon elliptic flow, even though the induced photon yields are marginal.

\section{Numerical simulations of weak magnetic photon emission}
\label{sec:num}


We now implement the dissipative corrections to 
the calculation of direct photons in realistic simulations based on event-by-event hydrodynamical modeling. In accordance with corrections from spatial gradients, the evolution of the medium should be described by viscous hydrodynamics, with shear and bulk viscous corrections. Analogously, in the presence of weak external electromagnetic fields, the corresponding dissipation appears not only in the phase space distribution, but also the characterization of the medium evolution. 

When the external electromagnetic fields are sufficiently weak, as the case we are considering for the evolving QGP in high-energy heavy-ion collisions, and in particular when $|eF^{\mu\nu}|\ll T\nabla \sim T^2$, 
conservation of energy and momentum $\partial_\mu T^{\mu\nu}=0$ is barely affected. Therefore, the bulk evolution of the background system should still be captured by the standard modeling based on viscous hydrodynamics. As a consequence, the charge independent observables, such as the collective flow and mean transverse momentum of hadrons which need not to be distinguished with respect to electric charges, should not be affected. 

There could be charge dependent signatures in QGP generated owing to the weak magentic field.  Even for a QGP which is locally charge neutral, 
weak external electromagnetic fields induce deviations between different charge components. For instance, let us consider a QGP medium with one flavor of quarks and anti-quarks, with their number density $n_+$ and $n_-$, respectively. Local charge neutrality condition requires that
\be
n_c = e Q_+(n_+ - n_-) = 0\,.
\ee
In each fluid cell, the external electromagnetic fields drives quarks and anti-quarks differently, so their velocities split,
\be
U_\pm^\mu = U^\mu \pm \Delta U^\mu\,.
\ee
By the assumption of weak field, the drift flow four velocity $\Delta U^\mu$ should be recognized as perturbations on top of the background hydrodynamic flow, $U^\mu$, which implies, $U^\mu \Delta U_\mu =0$.
These split velocities of quarks and aniti-quarks satisfy respectively the equation of motion, 
\be
\label{eq:duEM}
(e+P)D U^\mu_\pm = -\nabla^\mu P\pm Q_+ E^\mu n_\pm + O(\nabla^2)
\ee
where we have neglected higher order dissipative corrections. \Eq{eq:duEM} describes the acceleration of the charged components in a fluid cell due to gradient force as well as the electromagnetic forces. \Eq{eq:duEM} is equivalent to 
\be
\label{eq:duEM2}
(e+P)D U^\mu = -\nabla^\mu P + O(\nabla^2)\quad\mbox{and}\quad
(e+P)D \Delta U^\mu = \frac{1}{2}Q_+ E^\mu (n_++n_-) + O(\nabla^2)\,. 
\ee
While the first equation is the standard hydrodynamical equation of motion for the background neutral fluid, which gives rise to the fluid four velocity of the fluid cell, the second equation characterizes the development of velocity splitting. 

A non-relativistic version of \Eq{eq:duEM2} has been considered previously~\cite{Gursoy:2014aka}, 
assuming the balance condition between the gradient force and the electromagnetic force so that the QGP stays close to local thermal equilibrium. Accordingly, the splittings in the rapidity-odd direct flow $\Delta v_1$ between charged particles were recognized~\cite{STAR:2023jdd}. At the top energy of RHIC, the magnitudes of the splittings in direct flow of charged particles is found comparable to the background direct flow, i.e., $\Delta v_1 \sim v_1$. 
In fact, considering the fact that the electromagnetic force from the external magnetic field (as well as the external electrical field which we do not include in the current discussion) lies parallel to the direct flow, the balance of the electromagnetic force and gradient force must lead to $\Delta v_1 \sim v_1$.
Of course, splittings in higher order flows are more involved as the geometrical argument does not apply. 
As a consequence, as an ansatz, we should be allowed to simplify our simulations by assuming that the rapidity-odd dipole modes associated with $U^\mu$ and $\Delta U^\mu$  are comparable 
in this work, and we leave the consistent solution of $\Delta U^\mu$ for future analyses. In this way, the dipole mode can be captured correctly regarding the splittings in the charge dependent direct flow, which results in 
reliable descriptions of photon elliptic flow. However, since higher order modes are not fully characterized, the current formualtion is expected insufficient for higher order flow signatures of the direct photon.

\subsection{Event-by-Event Hydrodynamic simulations}

\begin{figure}
\begin{center}
\includegraphics[width=0.45\textwidth] {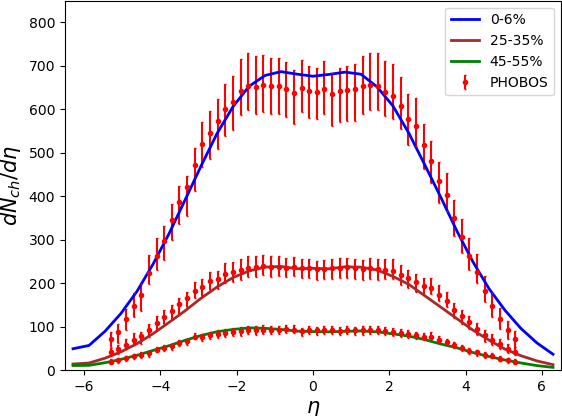}
\includegraphics[width=0.46\textwidth] {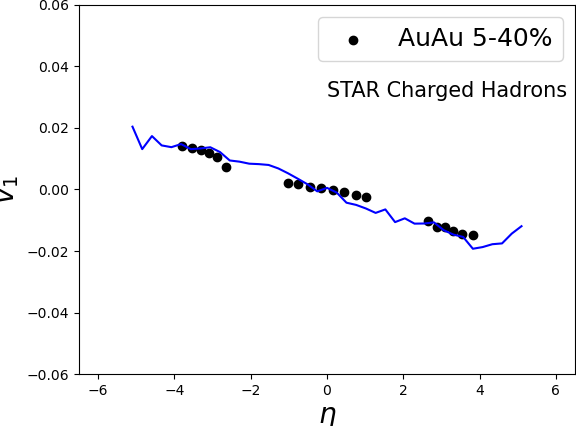}
\caption{
\label{fig:tunehydro}
Pseudo-rapidity dependent charged hadon yields (left panel) and charged hadron direct flow (right panel) at the top RHIC energy. Solid lines are corresponding results from event-by-event hydrodynamics simulations carried out in this work.
}
\end{center}
\end{figure}

\begin{figure}
\begin{center}
\includegraphics[width=1.\textwidth] {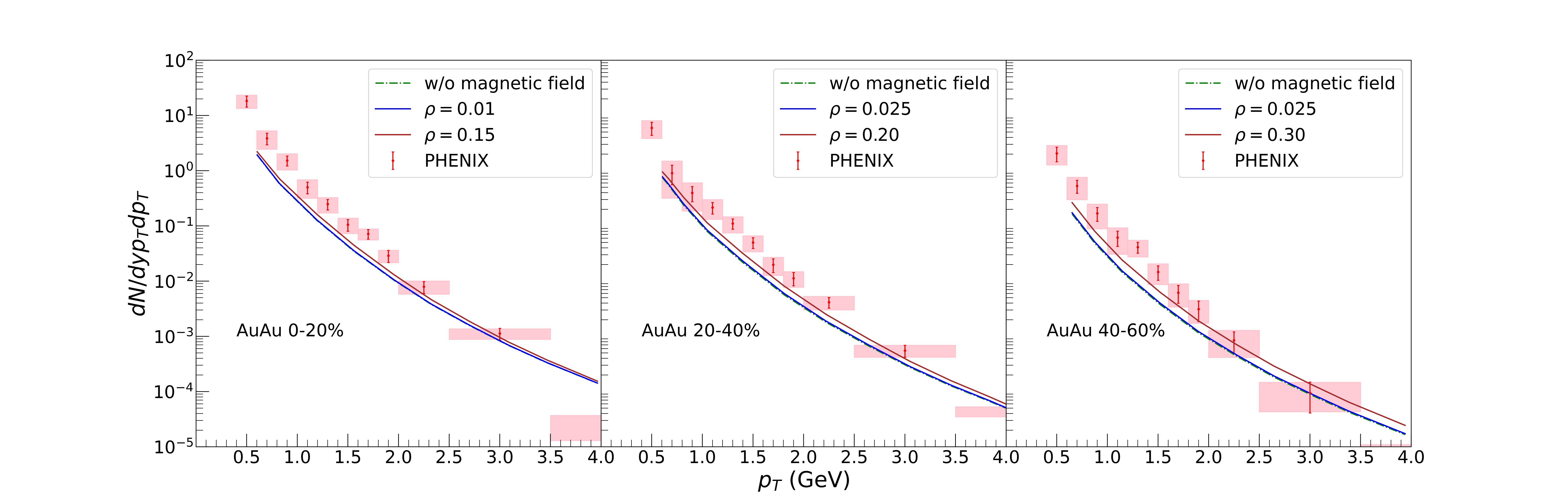}
\caption{
\label{fig:rhicv0}
Direct photon yields with also contributions from the weak magnetic photon emissions from AuAu collisions with $\sqrt{s_{NN}}=200$ GeV at RHIC~\cite{phenixPhotonv0}. Comparing to the background radiations, the weak magnetic fields lead to a small increase of photon yields. Given the values of $\rho$ that are compatible with the experimental data of $v_2^\gamma$, the increase ranges from less than 1\% to a few percents (0-20\% centrality), to about 10\% (20-40\% centrality) and to about 40\% (40-60\% centrality), respectively.
}
\end{center}
\end{figure}

In this current work, we take the existed results of thermal photon spectrum from the background medium from  \Reff{Gale:2021emg}. For the induced photon radiation from QGP by the weak external electromagnetic fields, we carry out event-by-event simulations of 3+1 dimensional viscous hydrodynamics. By doing so,  harmonic modes containing $\cos n\phi_p$ would arise automatically due to the initial geometrical fluctuations. Especially, the rapidity-odd dipole mode, namely, the mode associated with $\cos \phi_p$ 
with respect to a tilted fireball configuration~\cite{Chatterjee:2017ahy}, 
contributes to the photon elliptic flow~\cite{Sun:2023pil}. In a similar manner, the rapidity-odd quadrapole 
mode, namely, the mode associated with $\cos 2\phi_p$ with respect to a torqued fireball configuration~\cite{Bozek:2015bna}, contributes to the photon triangular flow. 

We solve viscous hydrodynamics using the state-of-the-art MUSIC program~\cite{Schenke:2010rr}, with respect to the 3+1 D T$_R$ENTo initial condition~\cite{Moreland:2014oya}, both of which have been applied extensively, for instance, by the JETSCAPE analyses~\cite{Putschke:2019yrg}. Following \Reff{PhysRevC.96.044912}, we apply a quite standard set of parameters to these simulations, so that the hadron spectra from experiments are well reproduced. These parameters include those characterizing the initial geometry of the system on an event-by-event basis, 
the ratio of shear viscosity to entropy density $\eta/s$ and a temperature dependent ratio of bulk viscosity to the entropy density $\zeta/s$. In particular, to be consistent with the previous calculations of thermal photon radiation from the background medium evolution, we choose the initial time at $\tau_0=0.4$ fm/c, and record all photons induced by the external electromagnetic fields up to the cross-over temperature $T_{c}=150$ MeV. 
As an illustration, from the event-by-event simulations, the resulted pseudo-rapidity dependent charged particle yields and direct flow $v_1$ are shown in \Fig{fig:tunehydro}, in comparison to the experimental data from RHIC~\cite{MultiExpStar,StarV1odd}. 

\Eq{eq:fem1b} 
requires the space-time information of the electromagnetic fields. 
Although the external electromagnetic fields can be well determined initially from the colliding nucleus, and are well orientated due to the collision geometry, 
how 
they evolve along with the medium expansion remains undetermined so far. In particular, the electric conductivity $\sigma_{\rm el}$ in QGP, which plays an essential role in the evolution of the electromagnetic fields, has large theoretical uncertainties.  
For instance, there can be order of magnitudes difference in $\sigma_{\rm el}/T$, depending on whether the QGP is strongly coupled or weakly coupled~\cite{Arnold:2000dr,Greif:2018ubz,Aarts:2020dda,Floerchinger:2021xhb}.
In this work, we focus on the effect of magnetic fields, and consider in the lab frame of the nucleus-nucleus collisions with only a non-zero y-component $\vec B = B_y \hat y$.
To avoid uncertainties from $\sigma_{\rm el}$ and its temperature dependence, as well as the unknown time evolution of the magnetic field, etc., we introduce a dimensionless and constant parameter in our simulations.
\be
\rho \equiv \frac{\sigma_{\rm el}}{T} \frac{\overline{e B_y}}{m_\pi^2}\,,
\ee
where $\overline{eB_y}$ can be regarded as the time averaged magnitude of the external magnetic field during the QGP stage, 
in the center of the fireball. Similarly, with a constant $\rho$ in the simulations, the ratio $\sigma_{\rm el}/T$ can be also understood as a time averaged value.
In realistic simulations, we shall allow $\rho$ to vary in order to reproduce experimental observables. 
{Let us emphasize that, treating $\rho$ a constant parameter in simulations, one implicitly takes into account of the medium effect on the decay of the magnetic field. For instance, in the most optimistic scenario, a constant $eB$ could be due to a sufficiently large electrical conductivity in QGP. In general, no matter how the magnetic field evolves in time, it is the time averaged field strength.}
As a somewhat preliminary calculation, we ignore the dependence of the magnetic field in the transverse directions (directions perpendicular to the beam axis). The longitudinal profile of the magnetic field is dominantly governed by the colliding nuclei, 
{regardless of the generation of a QGP medium~\cite{Yan:2021zjc},} which can be deduced via the Lienard-Wiechert potential, viz., $eB_y(\tau, \eta_s)=\overline{eB_y}\Gamma(\eta_s)$ with a time independent and normalized profile,
\be
\Gamma(\eta_s) \propto 
\frac{1}{(b^2/4+\gamma^2 \tau_0^2 (\sinh \eta_s+v \cosh\eta_s)^2)^{3/2}}+
\frac{1}{( b^2/4+\gamma^2\tau_0^2 (\sinh\eta_s- v \cosh\eta_s)^2)^{3/2}}
\ee
and $\Gamma(\eta_s=0)=1$. In practice, we fix $\tau_0=0.4$ fm/c in the above equation. Here, the space-time rapidity $\eta_s=\tanh^{-1} (z/t)$. The impact parameter $b$ and the Lorentz factor $\gamma=1/\sqrt{1-v^2}$ are to be fixed according to the centrality and the center-of-mass energy of the nucleus-nucleus collisions, respectively.

\subsection{Direct photon yields, $v_2^\gamma$ and $v_3^\gamma$ from RHIC and the LHC}


We first present our numerical results on the direct photon yields from the AuAu collisions at $\sqrt{s_{NN}}=200$ GeV in \Fig{fig:rhicv0}, with respect to the experimental measurements from three different centrality classes from the PHENIX collaboration. 
To be consistent with experiments, we also collect photons from the same rapidity window, $y\in [-1,1]$.
 In comparison to the direct photon productions from the background medium (green dot-dashed lines), the external magnetic field induces a small extra contribution. As one would expect, the resulted enhancement in the yields is proportional to the field strength. For instance, in the centrality class 20-40\% and with photon transverse momentum $p_T=1$ GeV, as we choose $\rho$ from $0.025$ to $0.2$, the direct photon yields receive an increase from $1\%$ to $15\%$, respectively.  Note that the 1\% increase is barely seen in the difference between the gree dot-dashed line and the blue line in \Fig{fig:rhicv0}.

\begin{figure}
\begin{center}
\includegraphics[width=1.\textwidth] {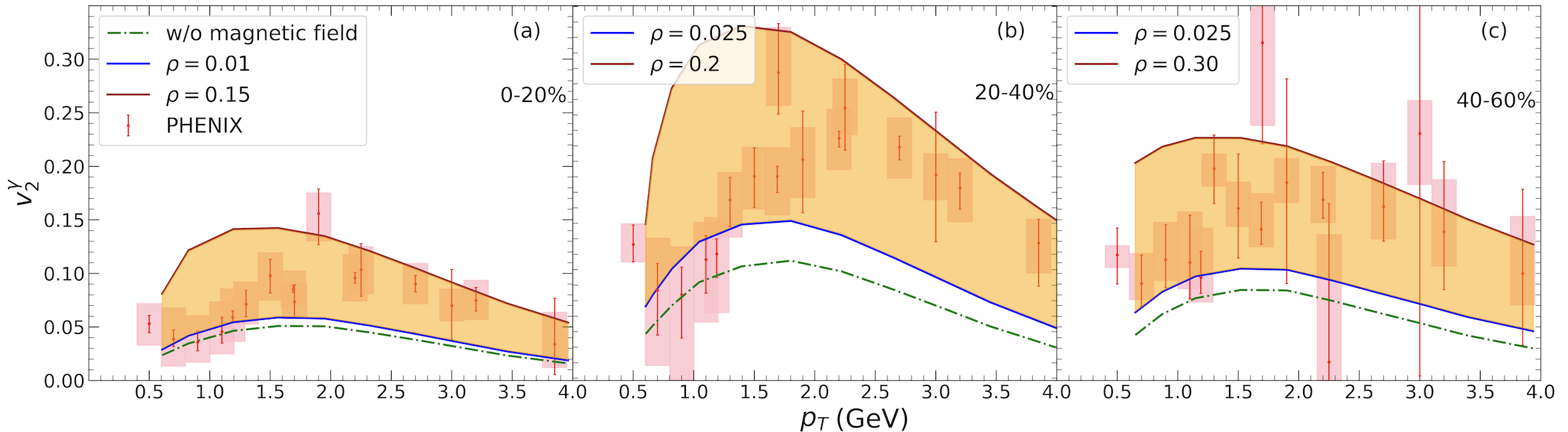}
\caption{
\label{fig:rhicv2}
Direct photon elliptic flow with also contributions from the weak magnetic photon emissions for three centrality classes at the RHIC with $\sqrt{s_{NN}}=200 $ GeV. In order to be compatible with the experimental data from the PHENIX collaboration~\cite{phenixPhotonv2}, theoretical results are shown as colored bands with respect to varying input values of parameter $\rho=\sigma_{\rm el}/T \times \overline{e B_y}/m_\pi^2$.
}
\end{center}
\end{figure}

Although it is insignificant to the yields, the magnetic field has a remarkable influence on the direct photon elliptic flow, as can be seen in \Fig{fig:rhicv2}. Again, as an example, one notices that in the centrality class 20-40\% and with photon transverse momentum $p_T=1$ GeV, $v_2^\gamma$ can rise from $30\%$ to a factor of 2.5, when $\rho$ is taken between $0.025$ and $0.2$. As a consequence of the sensitivity, one is allowed to use the experimentally measured direct photon elliptic flow to constrain the parameter $\rho$, which leads to the yellow band in \Fig{fig:rhicv2}. For all the given centrality classes, the experimental data are well described as the effect of magnetic field is included. Although the procedure of extracting $\rho$ is not accurate, and despite the large experimental errors, we find that the identified values of $\rho$ from $v_2^\gamma$ grow systematically as the centrality increases. This increase of $\rho$ is in qualitative agreement with theoretical expectations, that from central to peripheral collisions, there are more spectator nucleons in the colliding nucleus contributing to the generation of the external electromagnetic fields.

\begin{figure}
\begin{center}
\includegraphics[width=1.\textwidth] {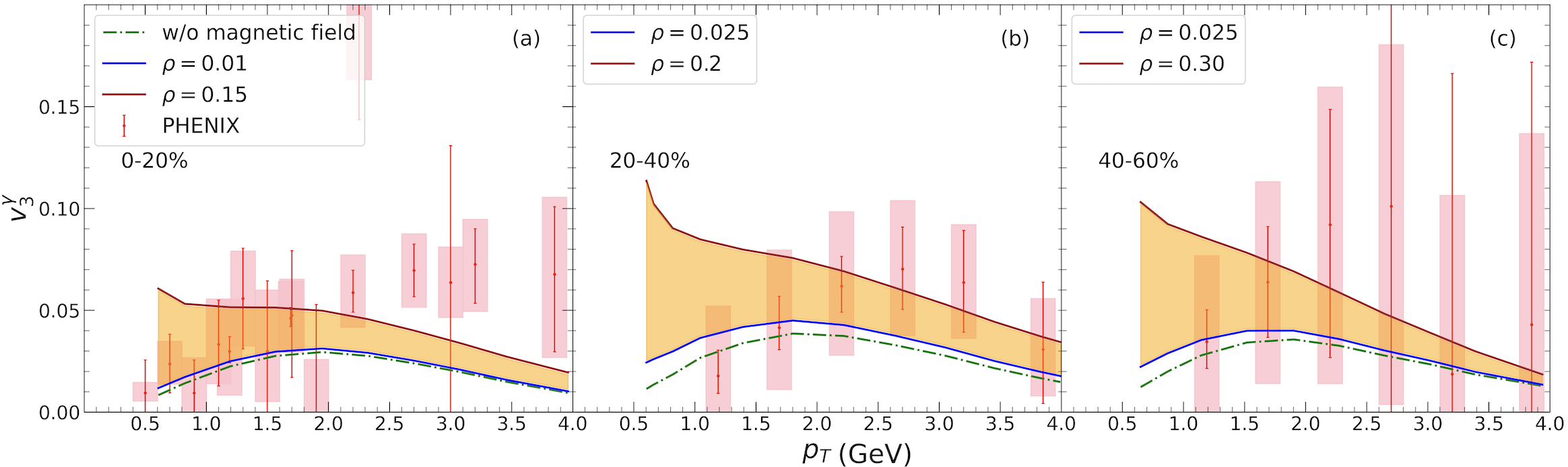}
\caption{
\label{fig:rhicv3}
Direct photon triangular flow with also contributions from the weak magnetic photon emissions for three centrality classes at the RHIC with $\sqrt{s_{NN}}=200 $ GeV, comparing to experimental data from the PHENIX collaboration~\cite{phenixPhotonv2}. Theoretical results are shown as colored bands with respect to input values of parameter $\rho=\sigma_{\rm el}/T \times \overline{e B_y}/m_\pi^2$ determined according to $v_2^\gamma$.
}
\end{center}
\end{figure}

With the same $\rho$ values identified with respect to the direct photon elliptic flow, we also calculate the triangular flow of the direct photons $v_3^\gamma$. In the previous work in \Reff{Sun:2023pil}, the initial medium density has been chosen from a smooth and tilted profile which does not lead to a rapidity-odd $\cos 2\phi$ mode in the expanding fireball, and as expected, triangular flow of photons cannot not be generated via the weak magnetic field. However, when geometrical fluctuations are considered on an event-by-event basis in a 3 dimensional system, rapidity-odd $\cos 2\phi$ mode emerges, we find a finite triangular flow contribution to the induced photons due to the weak magnetic field. 
As shown in \Fig{fig:rhicv3}, $v_3^\gamma$ of the direct photons gets increased when a larger external magnetic field is applied, yet it is less sensitive to the field strength comparing to $v_2^\gamma$. 

It should be emphasized that the triangular flow of the direct photons induced from a weak magnetic field is absolutely a novel effect. In the conventional mechanism of photon radiation involving a magnetic field, only even orders of harmonics in the photon spectrum appear according to the geometrical symmetry of the magnetic field. For instance, photons radiated via a synchrotron radiation has generically an elliptic flow, but no triangular flow. The generation of $v_3^\gamma$ from the weak magentic field demonstrates again the non-trivial interplay between the magnetic field and the longitudinal dynamics of the medium, with now the mode coupling involving the rapidity-odd $\cos 2\phi$ mode in the system expansion. 

As we have mentioned previously, the current framework cannot give rise to appropriate characterization of the higher order charge dependent harmonic modes in the fireball evolution. Consequently we do not expect good agreement of $v_3^\gamma$ comparing to experiment.

\begin{figure}
\begin{center}
\includegraphics[width=0.8\textwidth] {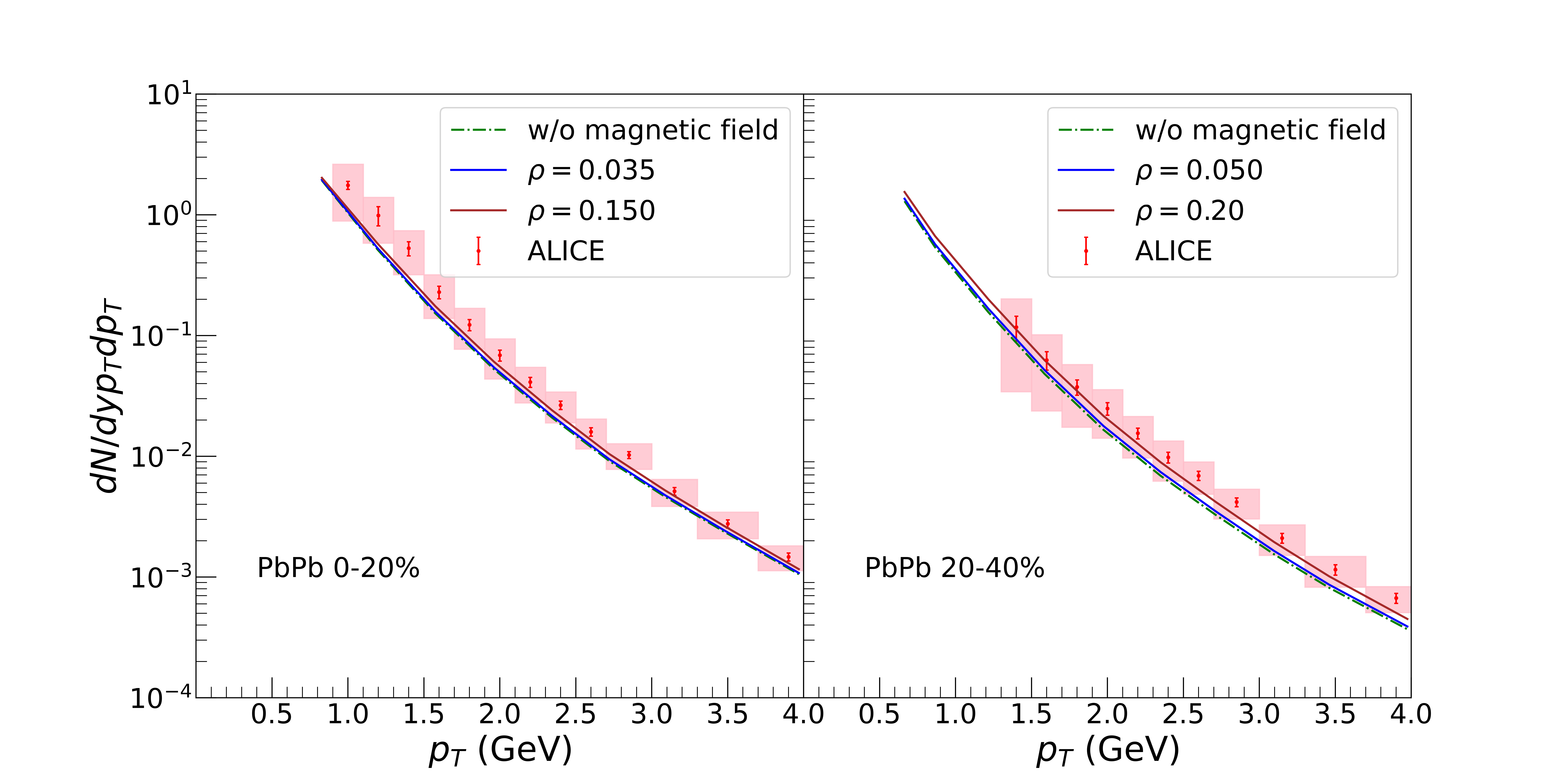}
\caption{
\label{fig:lhcv0}
Direct photon yields with also contributions from the weak magnetic photon emissions from PbPb collisions with $\sqrt{s_{NN}}=2760$ GeV at the LHC~\cite{AlicePhotonv0}. 
Given the values of $\rho$ that are compatible with the experimental data of $v_2^\gamma$, the increase ranges from less than 1\% to a few percent (0-20\% centrality), to about 10\% (20-40\% centrality) and to about 40\% (40-60\% centrality), respectively.
}
\end{center}
\end{figure}

\begin{figure}
\begin{center}
\includegraphics[width=.8\textwidth] {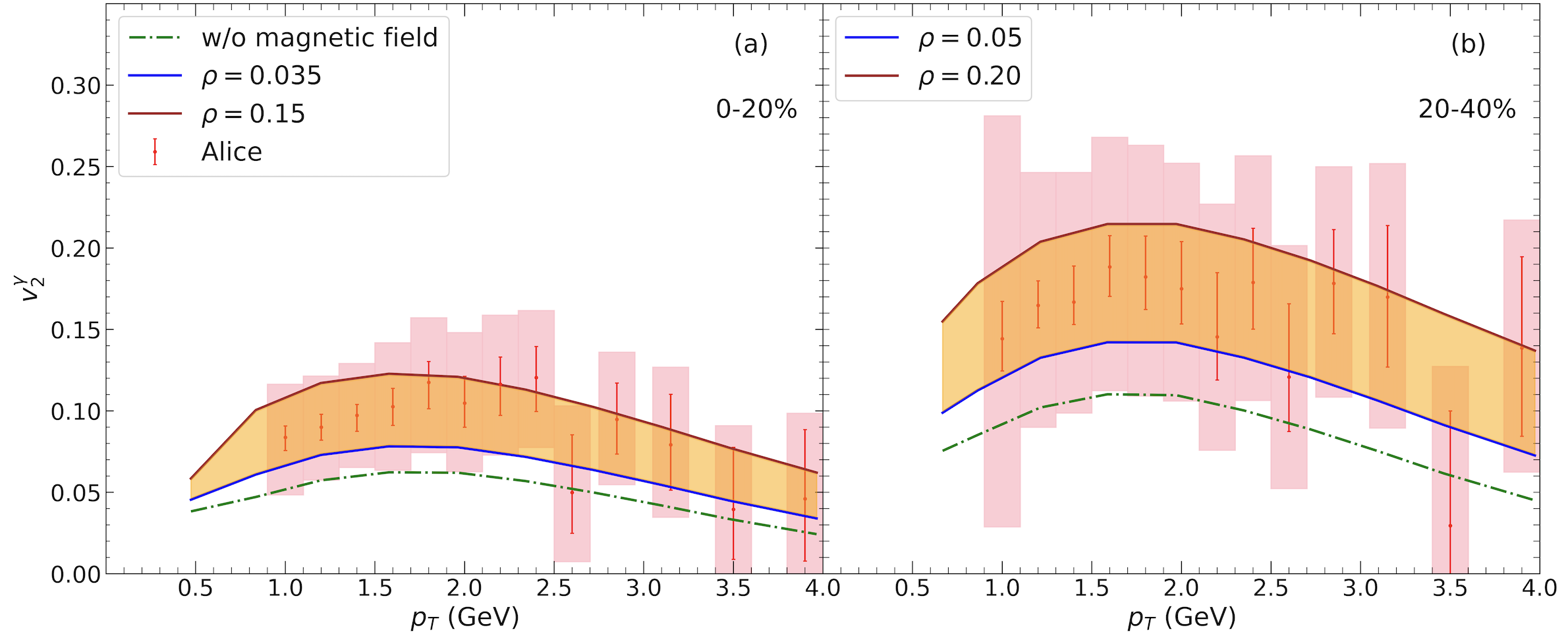}
\caption{
\label{fig:lhcv2}
Direct photon elliptic flow with also contributions from the weak magnetic photon emissions for two centrality classes at the LHC with $\sqrt{s_{NN}}=2760 $ GeV. In order to be compatible with the experimental data from the ALICE collaboration~\cite{ALICE:2018dti} considering only the statistical errors, theoretical results are shown as colored bands with respect to varying input values of parameter $\rho=\sigma_{\rm el}/T \times \overline{e B_y}/m_\pi^2$.
}
\end{center}
\end{figure}

The ALICE collaboration measured the yields and elliptic flow of direct photons from the PbPb collisions at the LHC, with $\sqrt{s_{NN}}=2760$ GeV~\cite{ALICE:2018dti}. Following the same strategy, we can in principle use the measured elliptic flow to constrain the values of $\rho$ in the our simulations. However, since the systematic uncertainties are too large, we only match the numerical solutions with respect to the data points with statistical errors. As shown in \Fig{fig:lhcv0} and \Fig{fig:lhcv2}, although the induced increase in the direct photon yields is marginal, the effect on the elliptic flow is remarkable.  For both centrality classes, the experimental data are well reproduced when the effect of a weak external magnetic field is included. Correspondingly, we notice a systematic growth in the values of $\rho$, from the central collisions (0-20\%) to the mid-central collisions (20-40\%).

%


\section{Extraction of the magnetic field from data}
\label{sec:bfield}

\begin{figure}
\begin{center}
\includegraphics[width=.5\textwidth] {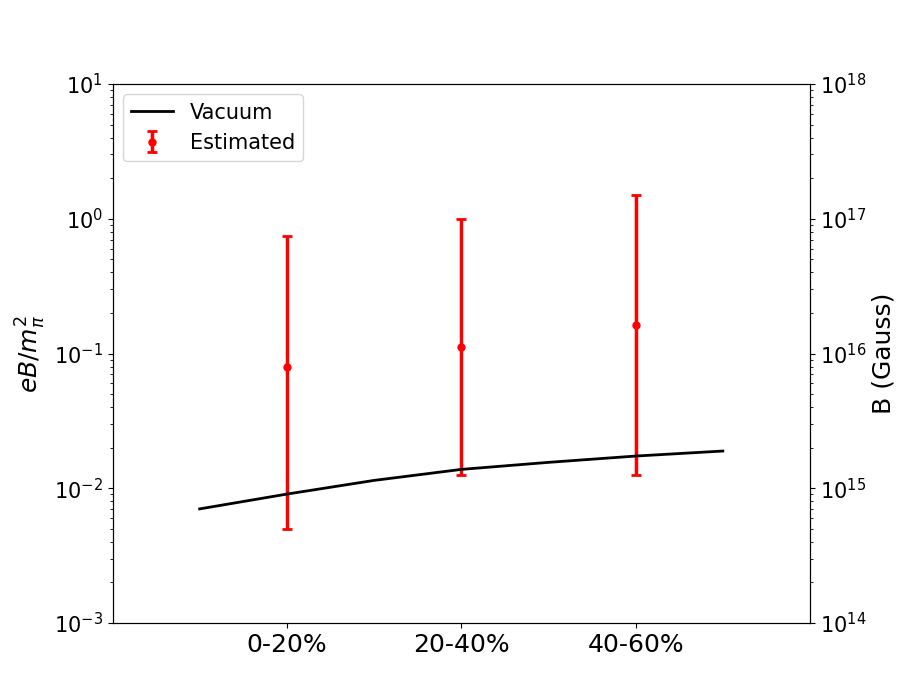}
\caption{
\label{fig:bvalue}
Extracted external magnetic field in RHIC AuAu collisions at $\sqrt{s_{NN}}=200 $ GeV, from the direct photon elliptic flow. The black line corresponds to the solution of the magnetic field evolution in vacuum in the center of the fireball at $\tau=0.4$ fm/c.
}
\end{center}
\end{figure}

We have extracted the dimensionless parameter $\rho$ from the measured data of $v_2^\gamma$ from the PHENIX collaboration, which allows us to further constrain the strength of the magnetic field. We take the ratio of electrical conductivity to temperature with respect to perturbative QCD calculation to the leading log order,  to be consistent with the photon production rate that has been considered~\cite{Gale:2021emg}, with $\sigma_{\rm el}/T\in [0.2,2]$~\cite{Floerchinger:2021xhb}. Given the theoretical uncertainties associated with the electrical conductivity, as well as the experimental error, the estimated values of the magnetic field for the corresponding three centralities are shown in \Fig{fig:bvalue}. In unit of pion mass square, the mean values of the magnetic fields in these collisions are found around 0.1. In context of QGP physics, these are indeed weak fields, especially one notices that the mean temperature square during the QGP evolution in heavy-ion experiments is $\langle T^2 \rangle \sim 4 m_\pi^2$.  
However, this extracted magnetic field in heavy-ion experiment is still rather strong in nature, considering, for instance, that the strongest magnetic field that has been deduced from the neutron star from X-ray spectrum is around $10^{13}$ Gauss~\cite{Kong:2022cbk}.

For comparisons, in \Fig{fig:bvalue} the theoretical prediction of the field strength is shown as the black solid line. This prediction is obtained by solving the vacuum time evolution of the external magnetic field, and evaluated at $\tau=0.4$ fm/c with respect to the center of the fireball, 
{
\be
B_y(\tau,z=0,\x_\perp=0) \propto  \frac{2 Z_{\rm eff}b}{[b^2/4+(\gamma v \tau)^2)]^{3/2}}
\ee
where $Z_{\rm eff}$ characterizes the electrical charge of the colliding nucleus.}
Namely, it is the theoretically expected initial value of the magnetic field as the QGP system starts to evolve hydrodynamically. One first notices a very similar centrality dependence of the extracted field strength (red points) and the theoretical expectation (black line), which is consistent with the fact that magnetic field should grow from central towards peripheral collisions, as more spectators are involved in the creation of the field.  
Despite the agreement within error bar, 
there exists apparent overestimate of the extracted field strength. The overestimate should be expected, at least for a couple of reasons. Firstly, the effect of electrical fields is neglected in the current framework. Secondly, the current calculation considers the magnetic field induced photon radiations above the cross-over temperature, while the similar mechanism which should also be applicable to the hadron gas has not been included. It is not surprising that both effects can lead to additional photon productions with large momentum anisotropy, hence reduce the extracted value of the field strength.

\section{Conclusion and discussion}
\label{sec:dis}

Through the event-by-event simulations of 3+1 dimensional hydrodynamics, we have demonstrated that a small dissipative correction due to the weak external magnetic field does induce an extra contribution to the thermal photon radiation from QGP. The weak magnetic photon emission is a novel effect that relies on the interplay of the QGP longitudinal expansion and the weak magnetic field. Although it is not of significance to the yields, the photons produced are highly anisotropic. 
With respect to the rapidity-odd dipole and quadrapole modes in the expanding QGP, elliptic flow and triangular flow of the photons can be generated in the presence of the weak magnetic field.  

In many perspectives, the current framework is still crude: The space-time profile of the magnetic field is simplifed. Dissipations from the magnetic forces correct only the quark distribution fuction, while their influences on the hydrodynamic equations of motion have been neglected. Effects from the electrical field have not been taken into account.
Nonetheless, with the weak magentic photon emission, 
we are able to provide the first {\it realistic} hydrodynamic modeling of photon production that successfully explains the observed direct photon momentum anisotropy, with only a weak external magnetic field that is consistent to theoretical expectations.
We therefore conclude that, the weak electromagnetic field, which must have been created in heavy-ion experiments, but has never been implemented in previous hydrodynamic calculations, is very likely responsible to the direct photon puzzle.
Accordingly, we propose that the direct photon elliptic flow can be taken as a probe to detect the magnetic field in heavy-ion experiments. In fact, it is a more sensitive probe to the magnetic field, than many other known signatures, such as the split between hyperon and anti-hyperon polarization~\cite{ALICE:2022wpn}. Moreover, direct photons detects QGP in the early stages, in which magnetic field is expected stronger.  
 

\acknowledgements

L.Y. is grateful for helpful discussions with G. Denicol and M. Luzum during the KITP workshop on ``The Many faces of Relativistic Fluid Dynamics''. 
This research was supported in part by the National Science Foundation under Grant No. NSF PHY-1748958, and 
by National Natural Science Foundation of China under Grant No. 11975079.

\appendix

\section{First order shear and bulk viscous correction $\delta f^{(1)}$}
\label{app1}

With respect to the symmetry of the tensor structure, the first order shear viscous correction in the distribution function is related to $\bra \nabla_\mu U_\nu \ket$. From the matching condition, one finds an integral equation according to the first term in \Eq{eq:df1},
\be
2\eta\bra \nabla_\mu U_\nu\ket  = -\frac{\tau_R }{T}\sum_a \int \frac{d^3 \p}{(2\pi)^3 E_p} P^\mu P^\nu P^\alpha P^\beta \bra \nabla_\alpha U_\beta\ket \frac{n_{a,{\rm eq}}'}{P\cdot U} \,.
\ee
For simplicity, we have neglected the species label in the momenta, but it should be aware that the operations of integration and summation are not commutable in the above equation.
The integral equation implies a solution $\eta = \tau_R \chi_{\rm shear}$, 
with, 
\begin{align}
\label{eq:kappa}
\chi_{\rm shear} &= - \frac{1}{15T}\sum_a\int \frac{d^3 \p}{(2\pi)^3 E_p} \frac{n'_{a,{\rm eq}}}{P\cdot U} (P^\alpha P^\beta \Delta_{\alpha\beta})^2
\xrightarrow[m\to 0]{} \frac{e+\P}{15 c_s^2}\,.
\end{align}
We introduce $\chi_{\rm shear}$ as the effective susceptibility associated with the shear viscous correction. It can be understood by noticing that the relaxation time $\tau_R$ plays the role of momentum diffusion constant. It can be also understood in terms of the equal-time two-point correlations of fluctuations. For instance, from the equal-time two-point correlation of the distribution function, $\bbra \delta f_a(\x,\p) \delta f_b(\x',\p') \kket =- n_{\rm eq}' (2\pi)^3\delta^{(3)}(\p-\p)\delta_{ab}\delta^{(3)}(\x-\x')$~\cite{landau1980statistical}, for the equal-time two-point correlations of the 
energy-momentum tensor fluctuations, one has
\be
\bbra \delta T^{\mu\nu}(t, \x) \delta T^{\alpha\beta}(t, \x')\kket\equiv 2 T  \Delta^{\mu\nu\alpha\beta} \chi_{\rm shear} \delta^{(3)}(\x - \x') + \ldots\,,
\ee
where the double brackets is used to indicate ensemble average over thermal fluctuations with respect to systems in local equilibrium. 
In the comformal limit, as shown in \Eq{eq:kappa}, $\chi_{\rm shear}$ reduces to a quantity proportional to the enthalpy density, and one recovers the well-known relation $\tau_R = 5 \eta/sT$.

The remaining of terms  in \Eq{eq:df1} all contribute to the bulk viscous correction to the distribution function. According to the matching condition, one obtains an equation,
\begin{align}
\label{eq:bulkeq}
-\zeta \nabla\cdot U
= &- \sum_a \int \frac{d^3 \p}{(2\pi)^3 E_p} P^\mu P^\nu  \frac{\tau_R}{P \cdot U}\frac{n_{a,{\rm eq}}'}{T}
\left[(P\cdot U)^2\left(c_s^2- \frac{1}{3}\right)+ \frac{m_a^2}{3}\right]\nabla\cdot U  \cr
&+ \sum_a \int \frac{d^3 \p}{(2\pi)^3 E_p} P^\mu P^\nu n_{a,{\rm eq}}'\left(\frac{P^\mu \delta U_\mu^*}{T} - \frac{P\cdot U}{T^2} \delta T^*\right)\cr
= & \tau_R T^4(\xi_1 \Delta^{\mu\nu} + \xi_2 U^\mu U^\nu) \nabla\cdot U - (\xi_3 \Delta^{\mu\nu} + \xi_4 U^\mu U^\nu) T^3 \delta T^*  + A^\mu \delta U^*_\mu\,,
\end{align}
where $\xi_i$'s and $A^\mu$ are dimensionless functions defined from various integrals involving the equilibrium distribution,
\begin{align}
\xi_1 &= - \frac{1}{3 T^5} \sum_a \int \frac{d^3 \p}{(2\pi)^3 E_p} (P^\alpha P^\beta \Delta_{\alpha\beta})\frac{n'_{a,{\rm eq}}}{(P\cdot U)}\left[(P\cdot U)^2\left(c_s^2- \frac{1}{3}\right)+ \frac{m_a^2}{3}\right] \,,\\
\xi_2 &= -\frac{1}{T^5}\sum_a \int \frac{d^3 \p}{(2\pi)^3 E_p} (P\cdot U)n'_{a,{\rm eq}}\left[(P\cdot U)^2\left(c_s^2- \frac{1}{3}\right)+ \frac{m_a^2}{3}\right] \,,\\
\xi_3 &= \frac{1}{3T^5} \sum_a \int \frac{d^3 \p}{(2\pi)^3 E_p} (P^\alpha P^\beta \Delta_{\alpha\beta}) P\cdot U  n'_{a,{\rm eq}}\,, \\
\xi_4 & =  \frac{1}{T^5} \sum_a \int \frac{d^3 \p}{(2\pi)^3 E_p}(P\cdot U)^3  n'_{a,{\rm eq}}\,.
\end{align}
In the massless limit, $m_a\to 0$, one finds, $\xi_1 =\xi_2/3$ and $\xi_3 = \xi_4/3$.
Comparing the tensor structure on both sides of \Eq{eq:bulkeq}, one recognizes the following solution to the equation,
\begin{subequations}
\begin{align}
\delta U^*_\mu  &= 0 \,,\\
\tau_R \xi_2 \nabla\cdot U &= \xi_4 \frac{\delta T^*}{T} \,, \\
\label{eq:zetadelta}
\tau_R T^4\left(\frac{\xi_3}{\xi_4} \xi_2 - \xi_1\right) &= \tau_R \chi_{\rm bulk}
= \zeta \,.
\end{align}
\end{subequations}
The last equation defines 
accordingly the effective susceptibility associated with the bulk viscous correction, 
\be
\chi_{\rm bulk} 
= T^4 \left(\frac{\xi_3}{\xi_4} \xi_2 - \xi_1\right) 
\ee


\bibliography{references}

\end{document}